\documentclass[a4paper,amsmath,amssymb,floatfix,footinbib,twocolumn,showpacs,preprintnumbers, subfigure]{revtex4-1}

\usepackage[FIGTOPCAP]{subfigure}
\usepackage{graphicx}    
\usepackage{dcolumn}    
\usepackage{bm}         
\usepackage{color}
\usepackage[colorlinks=true ,citecolor=blue]{hyperref}
\usepackage{amsmath,amsfonts}               
\newcommand{\ee}{\text{e}}
\newcommand{\ii}{\text{i}}
\newcommand{\so}{\text{SO}}
\newcommand{\nn}{\text{N}}

\begin{document}
\title{Dirac-Weyl electrons in a periodic spin-orbit potential}
\author{Lucia Lenz}
\author{Dario Bercioux}
\affiliation{  Freiburg Institute for Advanced Studies, Albert-Ludwigs-Universit\"at, D-79104 Freiburg, Germany\\
 Physikalisches Institut, Albert-Ludwigs-Universit\"at, D-79104 Freiburg, Germany}
 \pacs{61.48.De,72.80.Vp,72.25.Mk}
\begin{abstract}Graphene super-structures have been widely studied but the original form of the SU(2) Hamiltonian was never modified. We study SU(2)$\otimes$SU(2) super-structures arising from spatial modulation of spin-orbit interactions and derive an analytic band condition valid for a lattice momentum along the direction of modulation of the spin-orbit interactions. 
The simple form of this band condition enables us to estimate the size of gaps due to avoided band crossings and gives insight into the dependence of the band structure on the width of the potential. We also investigate band structures for the case where the lattice momentum forms a finite angle $\phi$ with respect to the modulation direction of the spin-orbit interactions.
\end{abstract}
\maketitle
\section{Introduction} Graphene, a two-dimensional (2D) network of $sp^2$ carbon atoms forming a honeycomb lattice, is well known for its low-energy electronic excitations. These can be described as Dirac-Weyl quasi-particles, characterized by a linear spectrum and a chiral nature leading to many of the unusual properties of this new material~\cite{reviews}. In the present letter we present yet another unusual effect, arising from the special structure of the Rashba Hamiltonian for graphene. We impose a super-lattice of periodically modulated spin-orbit interactions (SOIs) on graphene. As SOIs interact with the spin degree of freedom, the original SU(2) structure of the graphene Hamiltonian is modified to a SU(2)$\otimes$SU(2) structure. For such super-lattices it is usually not possible to obtain a simple band equation, unlike in the SU(2) case, where analytic band equations were already obtained for a large number of different cases~\cite{mckellar:1987,Barbier:2008,Barbier:2009,park:2008,dellanna:2010}. Despite this, for a lattice momentum along the modulation direction of the SOI, we derive analytic band equations, which look very similar to those obtained for the easier SU(2) case.
%
%
\begin{figure}[!t]
	\centering
	\includegraphics[width=0.9\columnwidth]{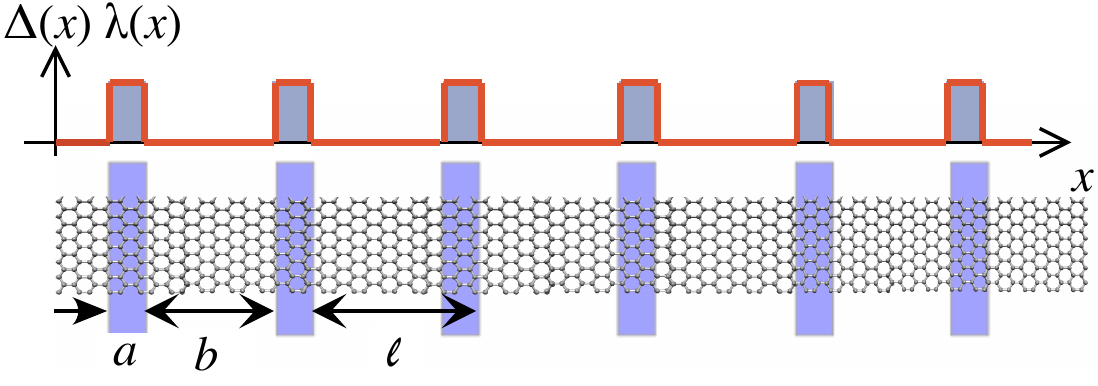}
	\caption{Sketch of the graphene plane in the presence of a modulation of the SOIs along the $x$-direction. The  periodicity length is $\ell=a+b$ with $a$ the length of the region with SOI. }\label{fig1}
\end{figure}
%
%

Graphene has two possible mechanisms of SOIs: the extrinsic and the intrinsic one. The \emph{intrinsic} SOI originates from carbon intra-atomic SOI. It opens a gap in the energy spectrum and converts graphene into a topological insulator with a quantized spin-Hall effect~\cite{kane:2005}.
This term has been estimated to be rather weak in clean flat graphene~\cite{soguinea,so2,so3,so4}. 
 However, a recent theoretical proposal shows that the presence of indium and thallium adatoms can enhance the gap associated to this effect. Even for low coverages of 6\%,  the gap for indium is of the order of 100~K, while for thallium it approaches room temperature~\cite{weeks:2011}.
The \emph{extrinsic} Rashba-like SOI originates  from interactions with the substrate, presence of a perpendicular external
electric field, or curvature of the graphene membrane~\cite{soguinea,so2,so3,so5,Konschuh:2010}. A series of recent experiments~\cite{varykhalov:2008,PhysRevLett.102.057602,varykhalov:2009} have reported the possibility to control and vary the Rashba term by opportune substrate engineering. A combination of the latter method and a nanomasking technique for controlling the adatom deposition would make variations of both the intrinsic and extrinsic SOI strengths feasible on sub-micrometer scales 
and thus our results applicable for experiments.

Modulated SOI leads to spin-polarized energy bands: this is essential for realizing graphene-based spintronics devices~\cite{spintronics} and especially for generating pure spin-currents, using, \emph{e.g.}, quantum pumping~\cite{spin:pumping} or the spin ratchet effect~\cite{Reimann:1997,smirnov:2008,scheid:2007}.

In the present work we introduce the spin-dependent transfer matrix (TM) method and use it to derive the spin-polarized energy bands in the case of SOI modulation. These simple, analytic band equations allow to directly obtain properties characterizing the band structure, say the band gaps and the dependence on the length of the SOI potential.

\section{Model and formalism}

We use the single valley Dirac-Weyl Hamiltonian~\cite{reviews}, which in absence of SOIs, reads
%
%
\begin{equation}\label{model}
 \mathcal{H}_0= v_{\text{F}} \bm{\sigma}\cdot\bm{p},
\end{equation}
%
%
where $v_{\text{F}}\approx 10^6$m s$^{-1}$ is the Fermi velocity, $\bm{\sigma}$ is the vector of Pauli matrices associated with the pseudo-spin and $\bm{p}=-\ii\hbar(\partial_x,\partial_y)$ the momentum operator in the 2D graphene plane. In the following we use units in which $\hbar=v_\text{F}=1$. 

As mentioned before two types of SOIs are allowed in graphene: the intrinsic and the extrinsic or Rashba SOI~\cite{kane:2005}.  The intrinsic SOI Hamiltonian is derived from a second nearest neighbor hopping term~\cite{soguinea,so5,Konschuh:2010}. In the low-energy limit the Hamiltonian reads 
%
%
\begin{equation}
  \mathcal{H}_\text{I}=\Delta(x)  \sigma_z s_z\,,
\end{equation}
%
%
where $\Delta(x)$ is a space dependent coupling constant, which can be implemented by appropriate deposition of adatoms. Here $s_z$ is the Pauli matrix associated with the spin of the Dirac-Weyl fermions. The intrinsic SOI does not lift spin degeneracy but opens a gap in the energy spectrum~\cite{kane:2005}. 

When inversion symmetry is broken, a Rashba SOI term is possible. It reads
%
%
\begin{equation}
  \mathcal{H}_\text{R}=\frac{\lambda(x)}{2}(\sigma_x s_y-\sigma_y s_x).
\end{equation}
%
%
Here, $\lambda(x)$ is a space dependent coupling. Contrary to $\mathcal{H}_\text{I}$ this term lifts spin degeneracy but does not open a gap~\cite{kane:2005,rashbagraphene}.

The single valley model~\eqref{model} is  valid within the approximation that the length scales of variation of the SOIs | via the $\lambda(x)$ and $\Delta(x)$ |  are assumed to be much larger than graphene's lattice constant $a_0\sim1.46$\AA\ but much smaller than the typical Fermi wavelength of quasiparticles $\lambda_\text{F}$. Since close to the Dirac points $\lambda_\text{F} \sim 1 / |\varepsilon|$, at low energy $\varepsilon$ this approximation is justified. This assumption ensures that the variation of SOIs, close to the $K$ points, can be approximated as a sharp change. Therefore we choose $\Delta(x)$ and $\lambda(x)$ to be periodic, piece-wise constant functions which are either zero or have magnitude $\Delta$ and $\lambda$, respectively (c.f. Fig.~\ref{fig1}).

In the following we introduce a generalization of the well known TM method~\cite{mckellar:1987}. It is valid when the lattice momentum is fixed along the modulation direction and leads to analytic band equations. For general directions of the lattice momentum, as discussed later on, we will consider a different method based on the Bloch theorem. 

%
%
\section{Spin-dependent TM method} 

We consider a unit cell of length $\ell=a+b$ with a region  $b$ without SOI and a region  $a$ with finite SOI as depicted in Fig.~\ref{fig1}. In the $n$th unit cell the wave function $\bm{\psi}(x)$ can be written as a superposition of the  forward (F) and backward (B) propagating spin up ($\uparrow$) and spin down ($\downarrow$) spinor solutions of the stationary Schr\"odinger equation $\Omega=\{\bm{\phi}_{\text{F},\uparrow},\bm{\phi}_{\text{B},\uparrow},\bm{\phi}_{\text{F},\downarrow},\bm{\phi}_{\text{B},\downarrow}\}$:
%
%
%
%
%
%
\begin{eqnarray*}
\bm{\psi}(x)= \bm{\mathcal{C}}_n\cdot \Omega(x-(n-1)\ell),
\end{eqnarray*}
%
%
where we have introduced the vectors of the coefficients in the $n$th unit cell $\bm{\mathcal{C}}_n=\{A_n,B_n,C_n,D_n\}$.
The TM $\mathcal{T}$ is the matrix describing the change of the amplitudes of $\bm{\psi}$ when moving between successive  unit cells:
%
%
%
%
\begin{equation}
\bm{\mathcal{C}}_{n+1}=\mathcal{T}\cdot
\bm{\mathcal{C}}_n.
\label{eq:transfer1}
\end{equation}
%
%
With $\Omega(x)$ being the matrix of the four component solutions, $\mathcal{T}$ can be written as~\cite{mckellar:1987}
%
%
\begin{equation}
 \mathcal{T}=\Omega(0)^{-1}\Omega(\ell).\label{eq:T}
\end{equation}
%
%
Since $\Omega$ obeys the Dirac equation, it has a constant determinant, therefore $\det[\mathcal{T}]=\det[\Omega(0)^{-1}\Omega(\ell)]=1$. Thus, the product of  eigenvalues  of $\mathcal{T}$ is one.
For a $4\times4$ TM, it is useful to specify how the four eigenvalues multiply to one. Because $\det[\Omega]$ is constant, we can fix 0 and $\ell$ in regions without SOIs,  where $\Omega$ is block diagonal with two sub-blocks which are themselves matrices built of two component spinors of $\mathcal{H}_0$. As the sub-block matrices are also obeying the Dirac equation,  their determinants are constant as well. Therefore the eigenvalues of $\mathcal{T}$ have to multiply to one pairwise. Together with the Born-von Karman requirement $\mathcal{T}^N=1$ | the eigenvalues of $\mathcal{T}$ must be the $N$th root of unity | the band condition can be written as
%
%
\begin{align}
 2\cos(\kappa \ell)+2\cos(\kappa^\prime \ell)=\text{Tr}[\mathcal{T}],\label{eq:trace}
\end{align}
%
%
where $\kappa$ and $\kappa'$ are the lattice momenta defined by $j2\pi/N\ell$ with $j$ integer.
In order to obtain the r.h.s. of this equation we need to know explicitly the matrices of spinors outside and inside the SOI regions, $\Omega_\text{N}$ and $\Omega_\text{SO}$, respectively. With these, Eq.~\eqref{eq:T} can be written as~\cite{mckellar:1987}
%
%
\begin{align*}
 \mathcal{T}=F_0^{-1} \Omega^{-1}_\nn(0)\Omega_\nn(l)\Omega^{-1}_\nn(a)\Omega_\so(a)\Omega^{-1}_\so(0)\Omega_\nn(0)F_0.
\end{align*}
%
Here, $F_0$ is a constant 4$\times$4 matrix describing the amplitudes in the $n$th unit cell. This matrix does not enter in the band conditions but cancels during the calculations. In the normal regions the matrix $\Omega_\nn$ is:
%
\begin{equation}\label{omega:N}
\Omega_\nn = \begin{pmatrix}
\text{sgn}(\varepsilon) \ee^{\ii k_x x} & -\text{sgn}(\varepsilon) \ee^{-\ii k_x x} \\
\ee^{\ii k_x x} \ee^{\ii \phi}& \ee^{-\ii k_x x}\ee^{-\ii \phi}
\end{pmatrix} \otimes \mathbb{I}_2,
\end{equation}
%
%
where $\phi=\text{arctan}(k_y/k_x)$ in the system without SOI. Here, $k_x,(k_y)$ is the wavevector outside the  SOI potential in $x,(y)$-direction and $\varepsilon$ the energy of the quasiparticles. Finally, $\mathbb{I}_2$ is the 2$\times$2 identity matrix in the spin subspace. Note, that the system is translational invariant along the $y$ direction. In the SOIs regions the matrix $\Omega_\so$ reads
%
%
\begin{equation}\label{omega:so}
\Omega_\so = \begin{pmatrix}
\ii \ee^{-\ii \xi_+} & \ii \ee^{\ii \xi_+} & -\ii \ee^{-\ii \xi_-} &  -\ii \ee^{\ii \xi_-} \\
\ii d_+ & -\ii d_+ & -\ii d_- & \ii d_- \\
d_+ & - d_+ &  d_- &  -d_- \\
\ee^{\ii \xi_+} & \ee^{-\ii \xi_+} & \ee^{\ii \xi_-} & \ee^{-\ii \xi_-}
\end{pmatrix} \cdot \mathcal{D}_0\,.
\end{equation}
%
%
In the following, $\alpha=\pm1$ refers to the two spin eigenmodes allowed by finite SOIs, and $\xi_\alpha=\text{arctan}(k_y/K_\alpha)$ are their angles of propagation with respect to the $x$-axis. The wavevectors $K_\alpha$ in $x$-direction are obtained by evaluating the eigenvalues of $ \mathcal{H}=\mathcal{H}_0+\mathcal{H}_\text{I}+\mathcal{H}_\text{R}$ for uniform SOIs and assuming energy conservation and translational invariance along the $y$ direction~\cite{bercioux:2010}. They are found to read
%
%
\begin{align}\label{k:alpha}
 K_\alpha(\varepsilon) &=\sqrt{\left(\varepsilon-\frac{\alpha\lambda}{2}\right)^2-\left(\Delta-\frac{\alpha\lambda}{2}\right)^2-k_y^2}\,.
\end{align}
%
%
Further, we have introduced the spin-dependent factors $d_\alpha= (\varepsilon-\Delta)/\sqrt{K_\alpha^2}|_{k_y=0}$, and the  diagonal matrix $\mathcal{D}_0=\text{diag}[\mathcal{N}_+,\mathcal{N}_+^*,\mathcal{N}_-,\mathcal{N}_-^*]$ with $\mathcal{N}_\alpha=\ee^{\ii K_\alpha x}/\sqrt{2(1+d_\alpha^2)}$.
Calculating the trace explicitly for $\phi=0$, the r.h.s. of Eq.~\eqref{eq:trace}  becomes
%
%
\begin{align}
 \text{Tr}[\mathcal{T}]=& \sum_{\alpha} \mathcal{F}_\alpha(k_x),
\label{eq:bandsum}
\end{align}
%
%
where we have introduced 
%
%
\begin{align} 
\mathcal{F}_\alpha(k_x)& =  -\left(\frac{K_\alpha}{\varepsilon-\Delta}+ \frac{\varepsilon-\Delta}{K_\alpha}\right) \sin(K_\alpha a)\sin(k_x b)\nonumber \\& +2 \cos(K_\alpha a)\cos(k_x b)\label{eq:falpha}.
\end{align}
%
%
Here, $K_\alpha$  and $k_x$ are the wavevectors inside and outside of the SOI regions, respectively. 
Note that the $\mathcal{F}_\alpha$ function does not depend on terms containing the opposite spin index. This comes from the fact that, if $\mathcal{T}$ were unitary, the trace would be the sum of two cosines. If $\phi=0$, all matrices building up $\mathcal{T}$ are unitary except for $\Omega_\so$. In $\Omega_\so$ the first two columns correspond to spinors associated with eigenvalues  $\alpha=+1$ while the second two columns correspond to spinors associated with the eigenvalue $\alpha=-1$. Therefore the first two spinors are orthogonal to the second two but only linearly independent to each other. Also the second two spinors are only linearly independent. This prevents $\mathcal{T}$ from being unitary and is the source of the factor in front of the sines in Eq.~\eqref{eq:falpha}. The special form of the four $2\times2$ subblocks of $\Omega_\so$ for $\phi=0$ thereby ensures that the sum of the two functions $\mathcal{F}_\alpha(k_x)$ exactly corresponds to the sum of the bands associated with the $\alpha=+1$ mode (plus-band) and the $\alpha=-1$ mode (minus-band)\cite{lenz:soon}. Hence, Eq. \eqref{eq:trace} can be written as two independent equations: 
%
%
\begin{align}
\begin{cases}
 2\cos(\kappa \ell)&  = \mathcal{F}_+(k_x) \\ 
2\cos(\kappa \ell) & = \mathcal{F}_-(k_x) 
\label{eq:bandfinal}
\end{cases}\,.
\end{align}
%
%
This spin-dependent TM method, which leads to a decoupling of the band equation \eqref{eq:trace} for the case $\phi=0$, represents the central analytical results of this work.
We have verified this result by comparing the bands calculated from Eq.~\eqref{eq:bandfinal} with those calculated with the method used for discussing finite $\phi$ in the last section of this letter [c.f. Eq.~\eqref{eq:bandcon}].

\section{Comparison with the dispersion relation for uniform SOI} 

%
%
\begin{figure*}[!th]
 \subfigure[][~$\lambda=1$ and $\Delta=1/4$] {\label{fig:bulkcomp-a}\includegraphics[width=0.32\textwidth]{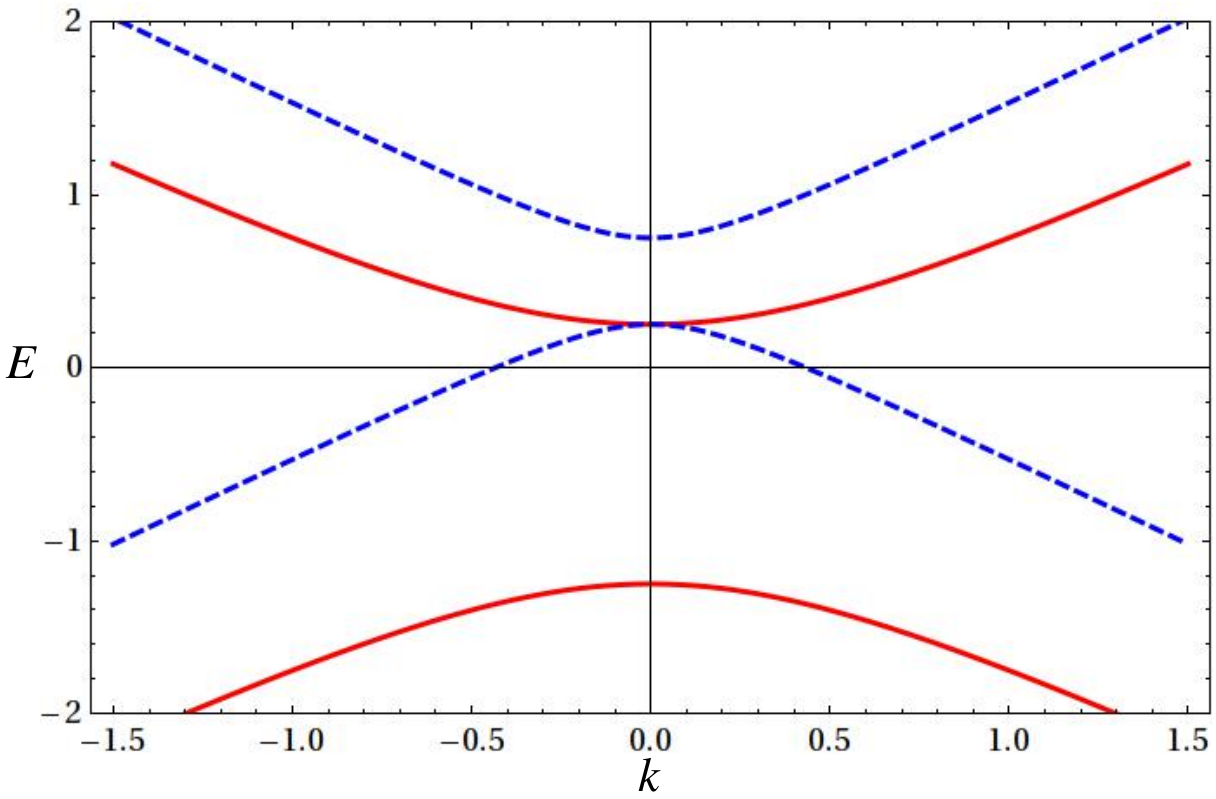}}
\subfigure[][~$\lambda=1/4$ and $\Delta=1/4$]{\label{fig:bulkcomp-b}\includegraphics[width=0.32\textwidth]{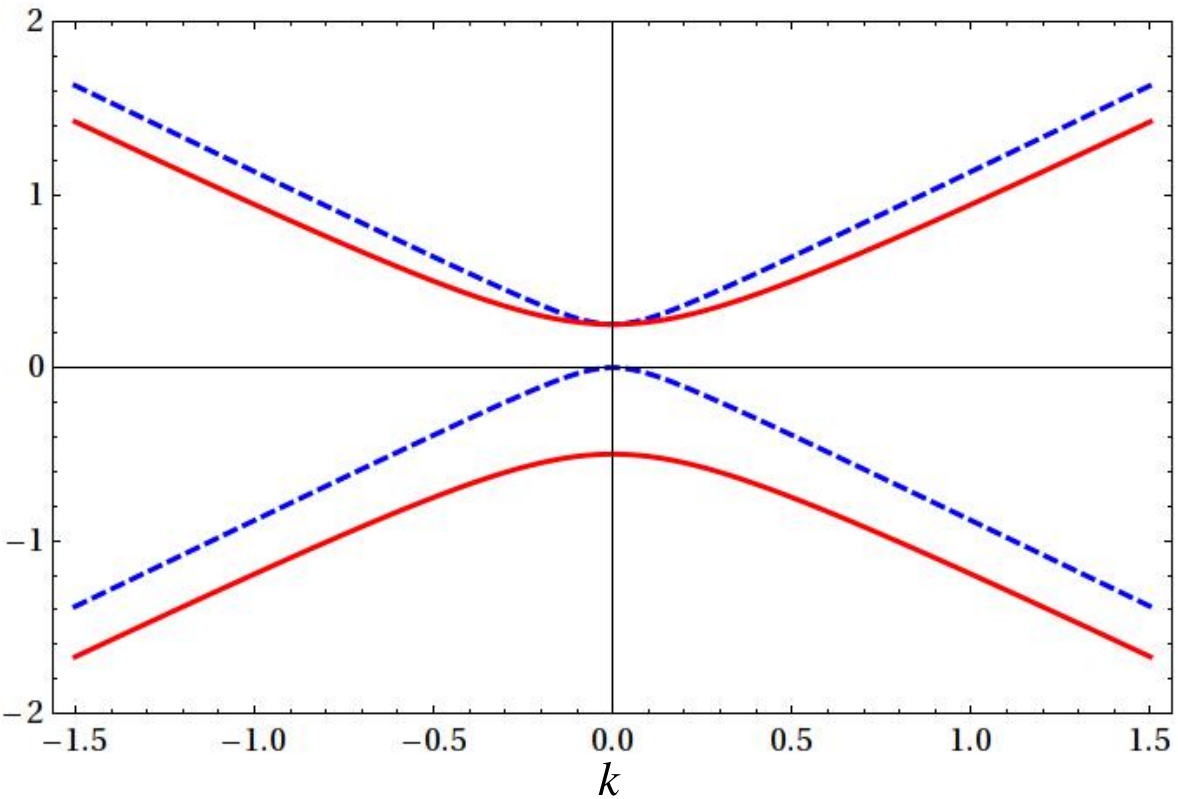}}
\subfigure[][~$\lambda=1/2$ and $\Delta=1/4$]{\label{fig:bulkcomp-c}\includegraphics[width=0.32\textwidth]{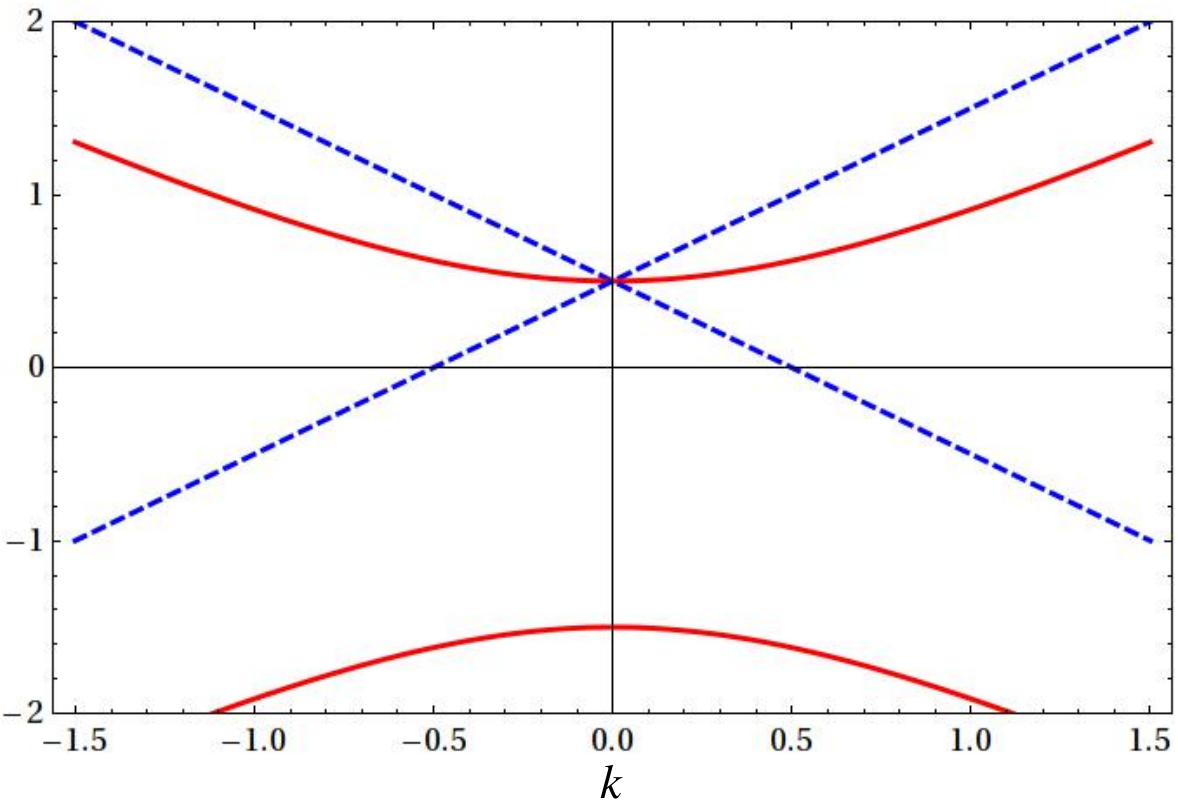}}
\subfigure[][~$\lambda=1$ and $\Delta=1/4$]  {\label{fig:bulkcomp-d} \includegraphics[width=0.32\textwidth]{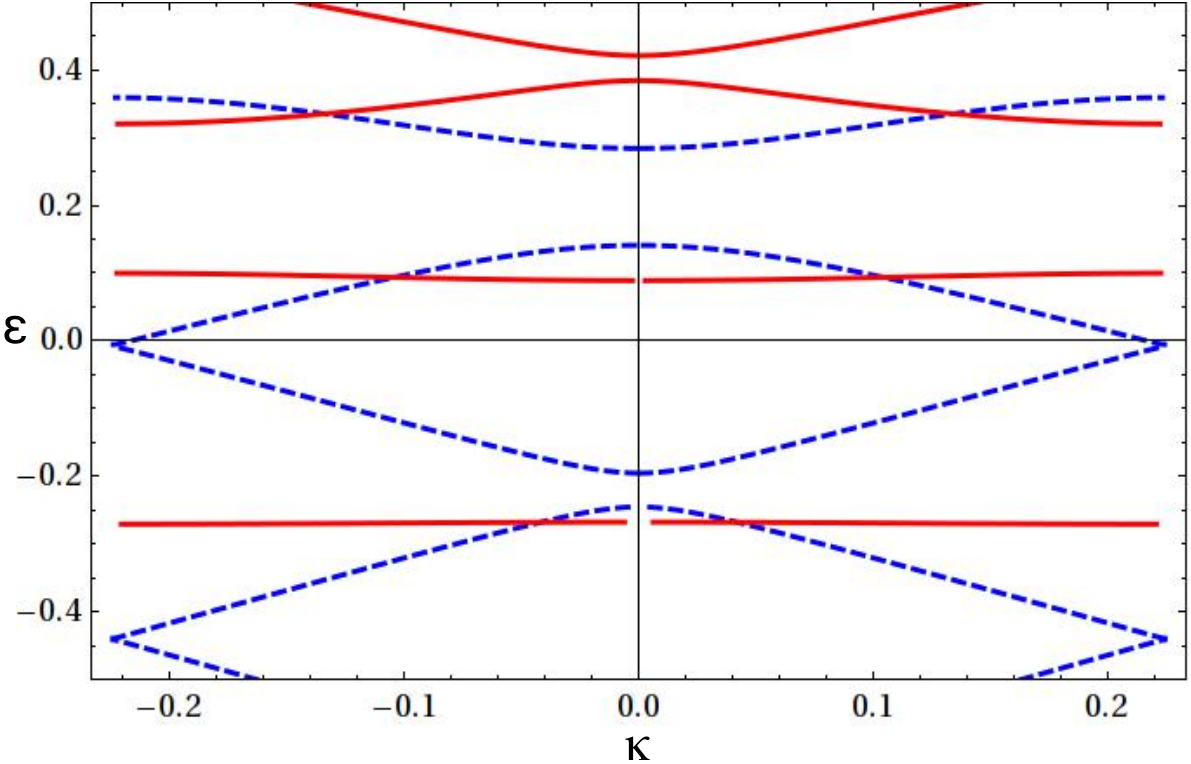}} 
\subfigure[][~$\lambda=1/4$ and $\Delta=1/4$]{\label{fig:bulkcomp-e} \includegraphics[width=0.32\textwidth]{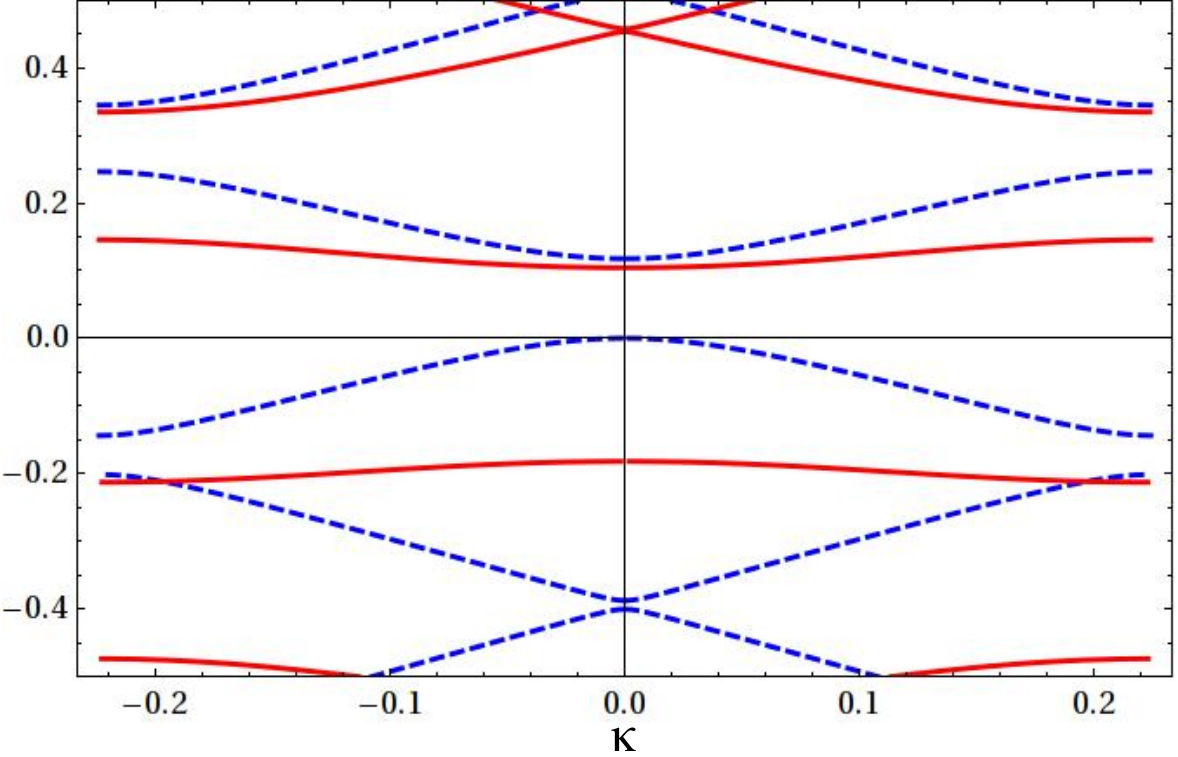}}  
\subfigure[][~$\lambda=1/2$ and $\Delta=1/4$]{\label{fig:bulkcomp-f}\includegraphics[width=0.32\textwidth]{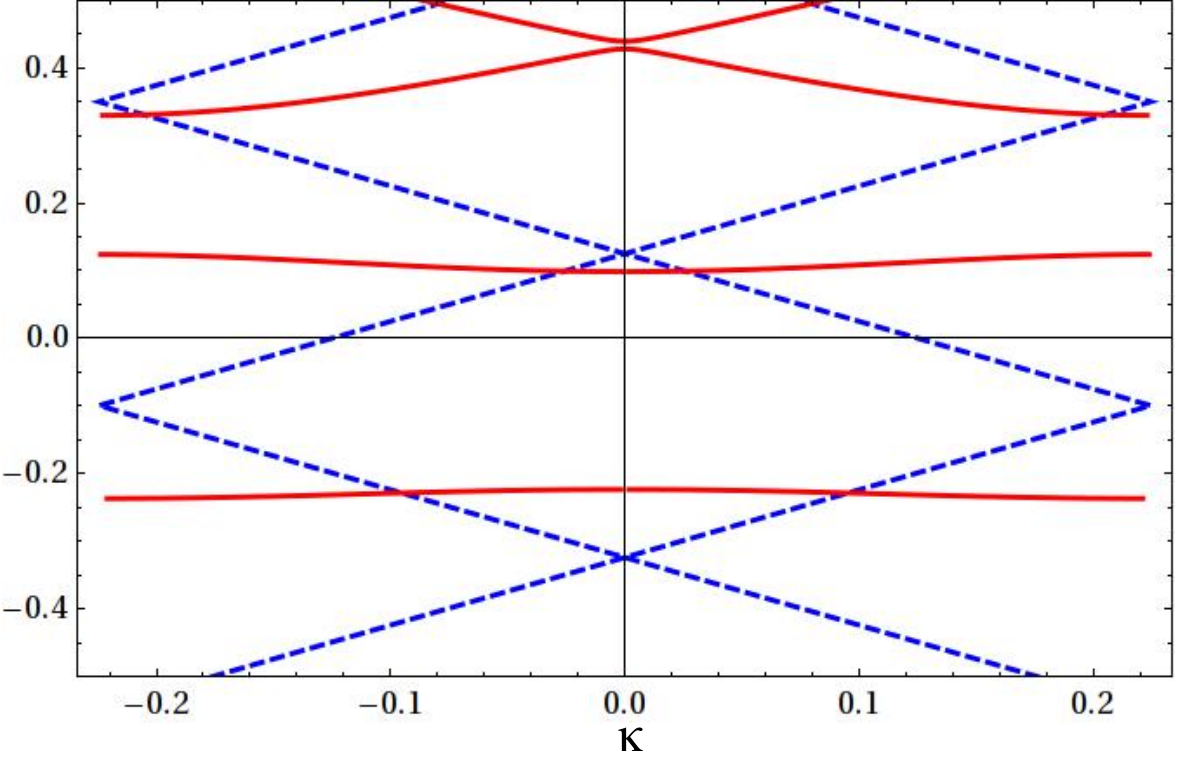}}
\caption{Panels (a)-(c) show the dispersion relation for uniform SOIs for different values of $\lambda$ and $\Delta$. Panels (d)-(f) show the dispersion relation for periodically modulated SOIs for the same $\lambda$ and $\Delta$ for comparison. The plus-bands (dashed-blue lines) are associated to the $\alpha=+1$ mode, the minus-bands (solid-red lines) to the $\alpha=-1$ mode. The  length of the SOI potential $a$ and the length without SOI $b$ are $a=b=7$.}\label{fig:bulkcomp}
\end{figure*}
%
%
 In order to understand what happens for systems with periodic SOIs, let us first review  what is known about graphene with uniform SOIs. There are three interesting regions in the $\lambda,\Delta$-parameter space:
(I) $\Delta/\lambda<1/2$, (II)  $\Delta/\lambda>1/2$ and
(III) $\Delta/\lambda=1/2$.

An example of the dispersion relations for uniform SOIs in these three regions are shown in Fig. \ref{fig:bulkcomp}\subref{fig:bulkcomp-a}--\subref{fig:bulkcomp-c}. The energy dispersions associated with $\alpha=+1$ ($\alpha=-1$) are depicted as dashed-blue (solid-red) lines. We denote the wave vector for the uniform case as $k$ and the energy in the uniform case as $E$ and shortly summarize the features for every case. Panel \subref{fig:bulkcomp-a} shows that for case (I) the spin degeneracy is completely lifted except for  $k=0$ where one of the valence and one of the conduction bands with opposite spin eigenstates touch. Panel \subref{fig:bulkcomp-b} shows case (II) in which the intrinsic SOI is dominating over the Rashba one. In this case an energy gap 
  opens  at $k=0$ where the plus and the minus conduction bands are degenerate. In both cases the overall behavior shows that the energy dispersions acquire a massive nature. In panel \subref{fig:bulkcomp-c} we show case (III). The energy bands of the plus spin state become degenerate at $k=0$ while for the minus spin state there is an energy gap.  
Importantly, for this case, the plus band keeps its linear dispersion, contrary to the minus band that still displays massive behavior. 

Now let us compare the dispersion relations for uniform SOIs to  band structures for modulated SOIs. The plus(minus)-bands associated with $\alpha=+1$ ($\alpha=-1$) are depicted as dashed-blue (solid-red) lines. In the following we use the Bloch wavevector $\kappa$ instead of $k$. In region (I), shown in panel \subref{fig:bulkcomp-d}, there is no gap between the valence plus band and the conduction minus band. Contrary to the case with uniform SOIs, the two bands cross | the valence band is shifted upwards in comparison to the conduction band. Therefore they now intersect and there is a degeneracy for two opposite and finite values of $\kappa$. Different is also that gaps | although small | open for both bands  at the Brillouin zone boundaries due to avoided band crossings, as it is usually the case when there is a periodic potential. The size of these gaps will be estimated later on. In panel \subref{fig:bulkcomp-e} we show region (II) with modulated SOIs. At $\kappa=0$ the  degeneracy of the two spin polarized conduction bands is lifted because of the different mass terms in the uniform SOI case. The plus valence band goes through $E=0$ and $k=0$ in the uniform case | therefore $K_+=0$ at this point and the plus valance band has to go through zero energy at  $\kappa=0$ in the periodic case as well. Again both bands have gaps at the Brillouin zone boundary. For region (III), shown in panel \subref{fig:bulkcomp-f}, this is not the case. Due to the linear dispersion of the plus band backscattering is forbidden and the band is gapless even at the zone boundaries~\cite{park:2008}. This can be seen in the band Eq.~\eqref{eq:bandfinal} as well. For $\Delta/\lambda=1/2$ the prefactor of the sines of the plus band is two and the band factorizes to $\cos(\kappa \ell)=\cos(K_+ a+k_x b)$. Also $K_+$ is a linear function of $\varepsilon$ for this case.
Concluding this comparison we can say that all general features that we find in the band structure for periodic SOIs at $\phi=0$ are understood from the dispersion relation of uniform SOIs.

\section{Band gaps at Brillouin zone boundary} 
As announced before, Eq.~\eqref{eq:bandfinal} enables us to estimate the band gaps at the Brillouin zone boundary. Always when the expression on the left hand side exceeds modulus two there is a gap in the energy spectrum since the equation has no real valued solution. In order to calculate an approximation for the size of the gaps, we follow a perturbative approach on the basis of  Ref.~\cite{ashcroft:book}. First we consider $\lambda=0$ and $\Delta\ll1$. The gap in this case is $2|U_{K=0}|$, where $U_{K=0}=-\ii \Delta/\pi$ is the leading component of the Fourier expansion of the piece-wise constant function modulating the intrinsic SOI. For simplicity we assume $a=b$ but we obtain a similar solution for $a\not=b$. The band gap for the case without Rashba SOI is thus approximately $2 \Delta/\pi$. When $\lambda\not=0$ we can substitute $\tilde{\Delta}=\Delta-\alpha\lambda/2$ and $\tilde{\varepsilon}=\varepsilon-\alpha\lambda/2$ in Eq.~\eqref{eq:bandfinal}. Up to first order in the expansion parameter $b\lambda$ this becomes
%
%
\begin{align}
2\cos(\kappa \ell)= & 2\cos(\tilde{K}_\alpha a)\cos(\tilde{\varepsilon} b)\\
& \hspace{-1cm}-\left(\frac{\tilde{K}_\alpha}{\tilde{\varepsilon}-\tilde{\Delta}}+ \frac{\tilde{\varepsilon}-\tilde{\Delta}}{\tilde{K}_\alpha}\right) \sin(\tilde{K}_+ a) \sin(\tilde{\varepsilon}b)\notag\\
& \hspace{-1cm}-\alpha \lambda b\left[\sin(b \tilde{\varepsilon})\cos(\tilde{K}_\alpha a)\right. \notag \\
& \hspace{-1cm} + \left(\frac{\tilde{K}_\alpha}{\tilde{\varepsilon}-\tilde{\Delta}}+ \frac{\tilde{\varepsilon}-\tilde{\Delta}}{\tilde{K}_\alpha}\right)\sin(\tilde{K}_\alpha a)\cos(b \tilde{\varepsilon})\Big] + \mathcal{O}(b\lambda)^2 \notag,
\end{align}
%
%
with $K_\alpha(\tilde{\varepsilon},\tilde{\Delta})\equiv\tilde{K}_\alpha$. 
The first order term for $b\lambda\ll1$ is rather small and it will never lead to the opening of gaps. Actually, when the zeroth order term approaches an extremum as a function of $\tilde{\varepsilon}$, the first order term approaches zero.  Thus it is a reasonable approximation to neglect the first order term. Therefore | in the perturbative approximation | the gaps of the bands for $\lambda\not=0$ are
%
%
\begin{align}
\varepsilon_{\text{gap}}\approx\frac{2 (\Delta-\alpha\lambda/2)}{\pi}.\label{eq:gaps}
\end{align}
%
%
These values coincide with the numerical results.
%
%
\begin{figure}[!t]
\centering
\includegraphics[width=0.9\columnwidth]{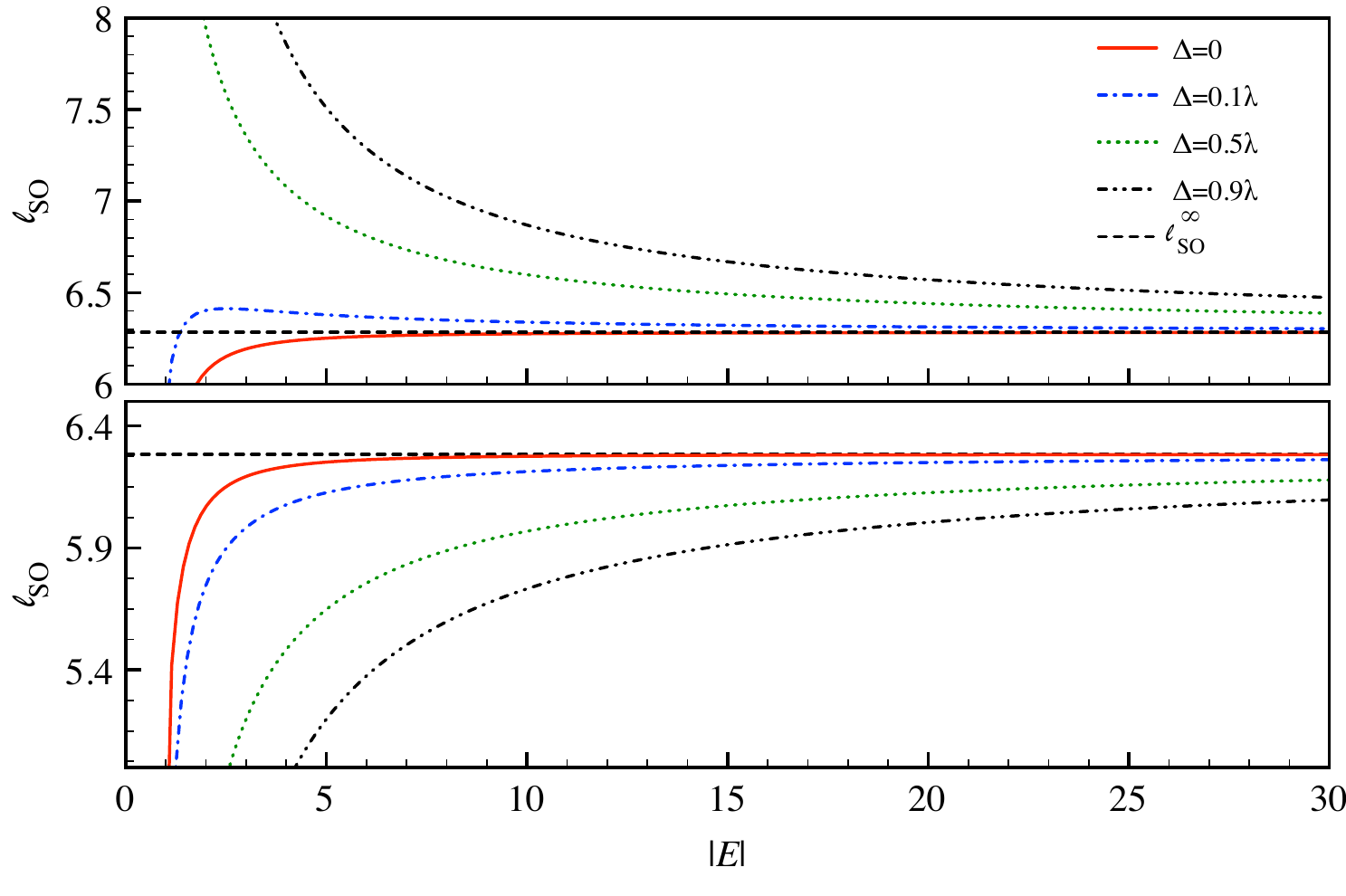}\caption{The dependence of $\ell_\so$ on $E>0$ (upper panel) and $E<0$ (lower panel) for $\lambda=1$ and different $\Delta$ is shown. The black dashed line refers to the approximation of $\ell_\so$ for large $E$, c.f.~Eq.~\eqref{eq:spinprec}. }\label{fig:lsoplot}
\end{figure}
%
%
%
%
%

\begin{figure}
\centering
\subfigure[][~$a=\ell_\so^\infty$]{\label{fig:spinprec-a}\includegraphics[width=0.48\columnwidth]{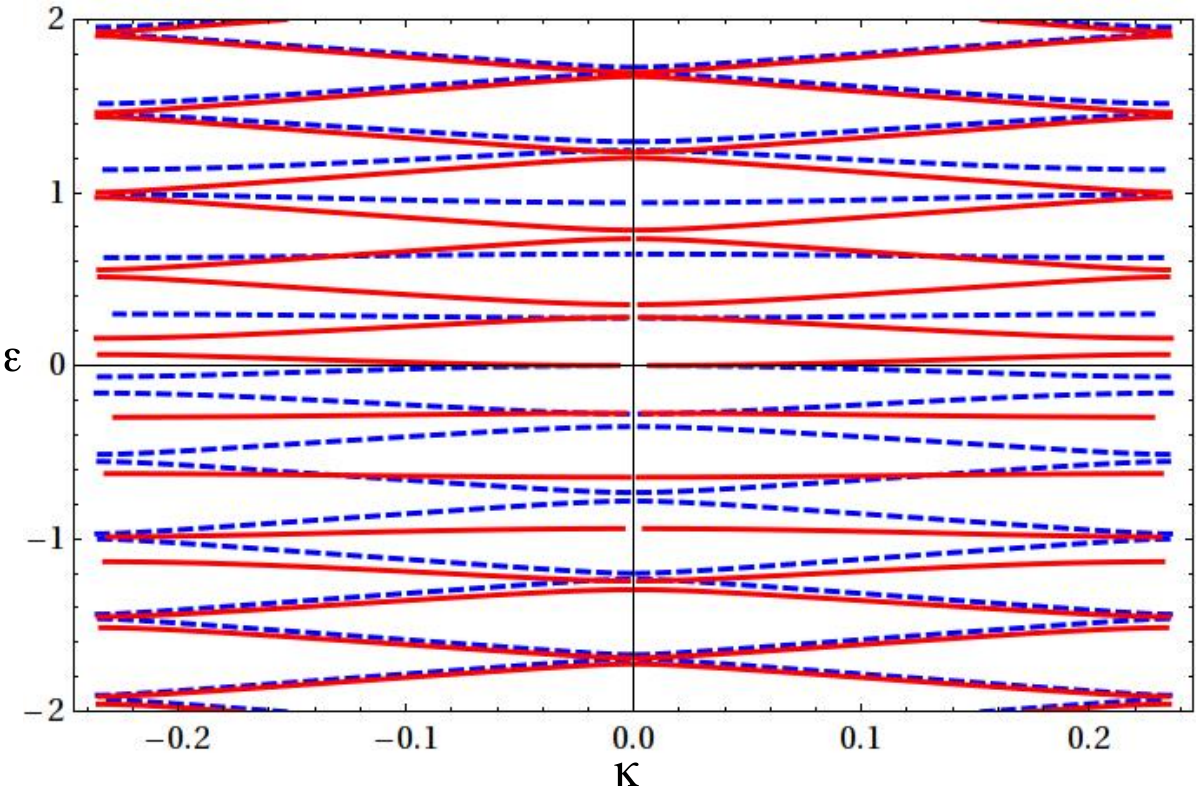}} \hspace{0.1cm}
\subfigure[][~$a=\ell_\so^\infty/2$]{\label{fig:spinprec-b}\includegraphics[width=0.48\columnwidth]{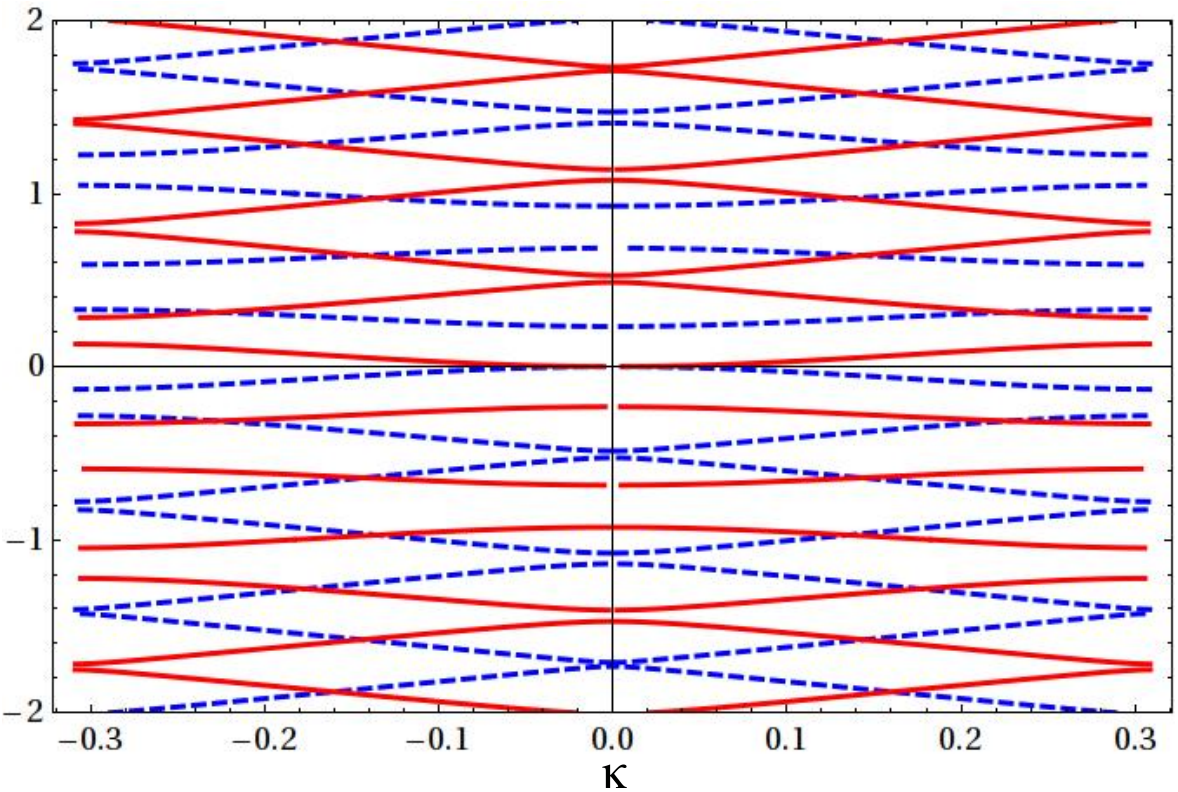}}
\caption{Band structure for $\lambda=1$ and $\Delta=0$ and $b=7$ for different lengths of the SO potential. Panel (a):  $a=2\pi/\lambda=\ell_\text{SO}^\infty$ and panel (b): $a=\pi/\lambda=\ell_\text{SO}^\infty/2$.The dashed-blue lines correspond to the plus-band and the solid-red lines to the minus-band. }\label{fig:spinprec}
\end{figure}

%
%
\section{Dependence  on the spin-precession length}

Finally we want to examine how the band structure depends on the periodicity length of the SOIs. As noticed in \cite{bercioux:2010} the transmission through a SOI barrier depends on the length of the barrier compared to the spin precession length and so does the band structure for periodic SOIs. First we derive the spin precession length $\ell_\so$,  which is the length that a carrier  needs to cover in order to go back to the initial spin state. In the ballistic limit the spin-precession length is expressed by $\ell_\so=2\pi/(K_--K_+)$.
A plot of $\ell_\so$ as a function of $E$ is shown in Fig.~\ref{fig:lsoplot}. Note that the behavior of $\ell_\so$ is strongly dependent on $\Delta$ but the asymptotic value for large $E$ is not. Therefore we expand $K_\alpha$ in the limit $E\gg\lambda,\Delta$. This yields a spin precession length independent of $\Delta$ and $E$:
%
%
\begin{align}
\ell_\so^\infty\approx\frac{2\pi}{\lambda}\,.
\label{eq:spinprec}
\end{align}

In Fig.~\ref{fig:spinprec} the band structure for $a=\ell_\so^\infty$ and the band structure for $a=\ell_\so^\infty/2$ is shown in panel \subref{fig:spinprec-a} and \subref{fig:spinprec-b} respectively. In the first case the plus and the minus bands lie on top of each other for large energies, where the approximation within $\ell_\so^\infty$ was calculated is valid. This is easily verified by expanding Eq.~\eqref{eq:bandfinal} in the same limit up to first order. One can show that if $a=\ell_\so^\infty$ for large energies only the pre-factor of the sines in Eq. \eqref{eq:bandfinal}  depends on $\alpha$. The pre-factor becomes closer and closer to one for increasing $\varepsilon$. When $a=\ell_\so/2$ a similar analysis shows that the band equation \eqref{eq:bandfinal} depends on $\alpha$ | in this case $\alpha$ is the overall sign | and the two bands are shifted with $\pi$ in their oscillation.
%
%
%
\begin{figure*}[!th]
\subfigure[][~$\lambda=1$ and $\Delta=0$]{\label{fig:contourplot-b}\includegraphics[width=0.235\textwidth]{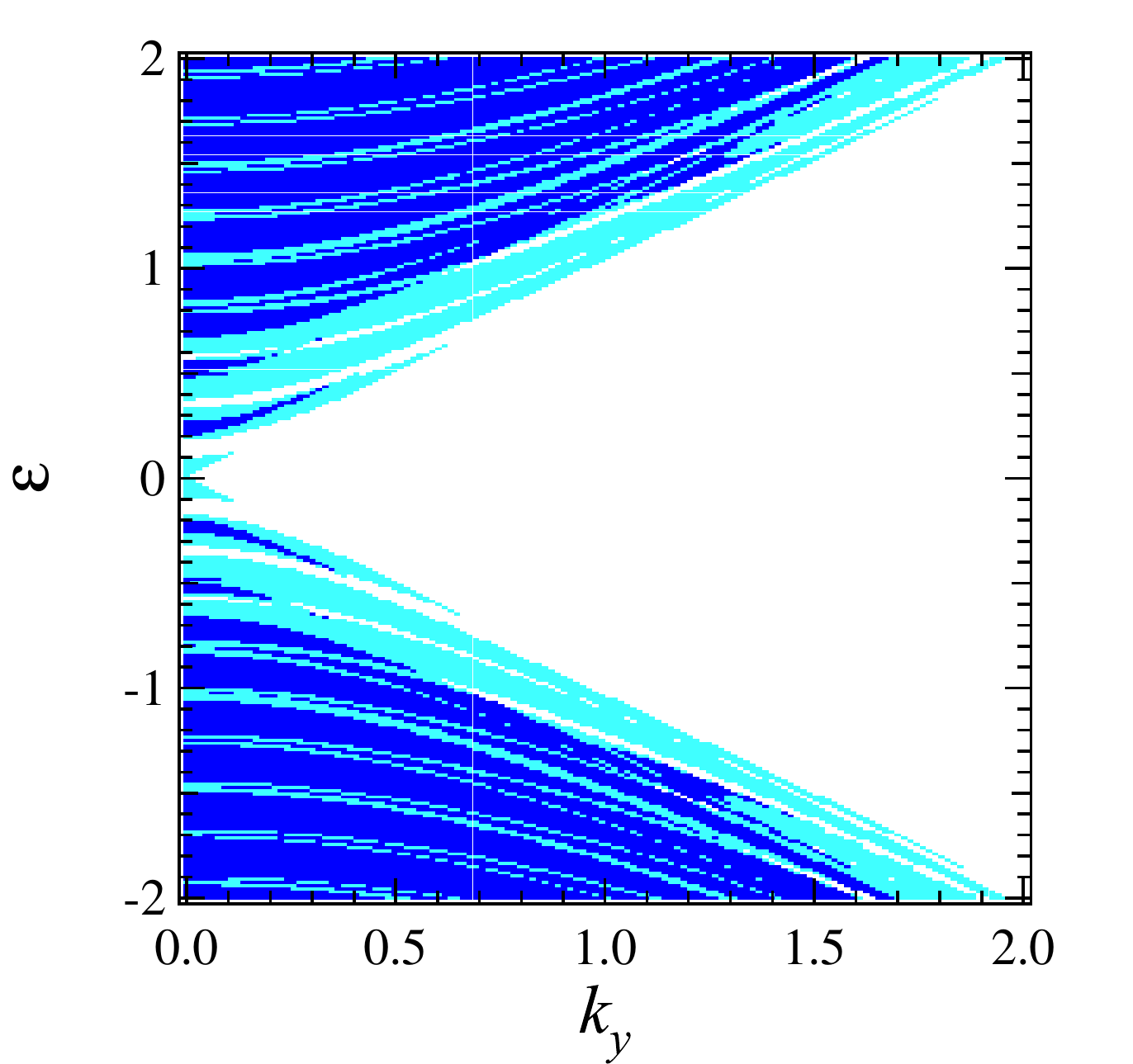}}
\subfigure[][~ $\lambda=0$ and $\Delta=1/2$]{\label{fig:contourplot-a}\includegraphics[width=0.23\textwidth]{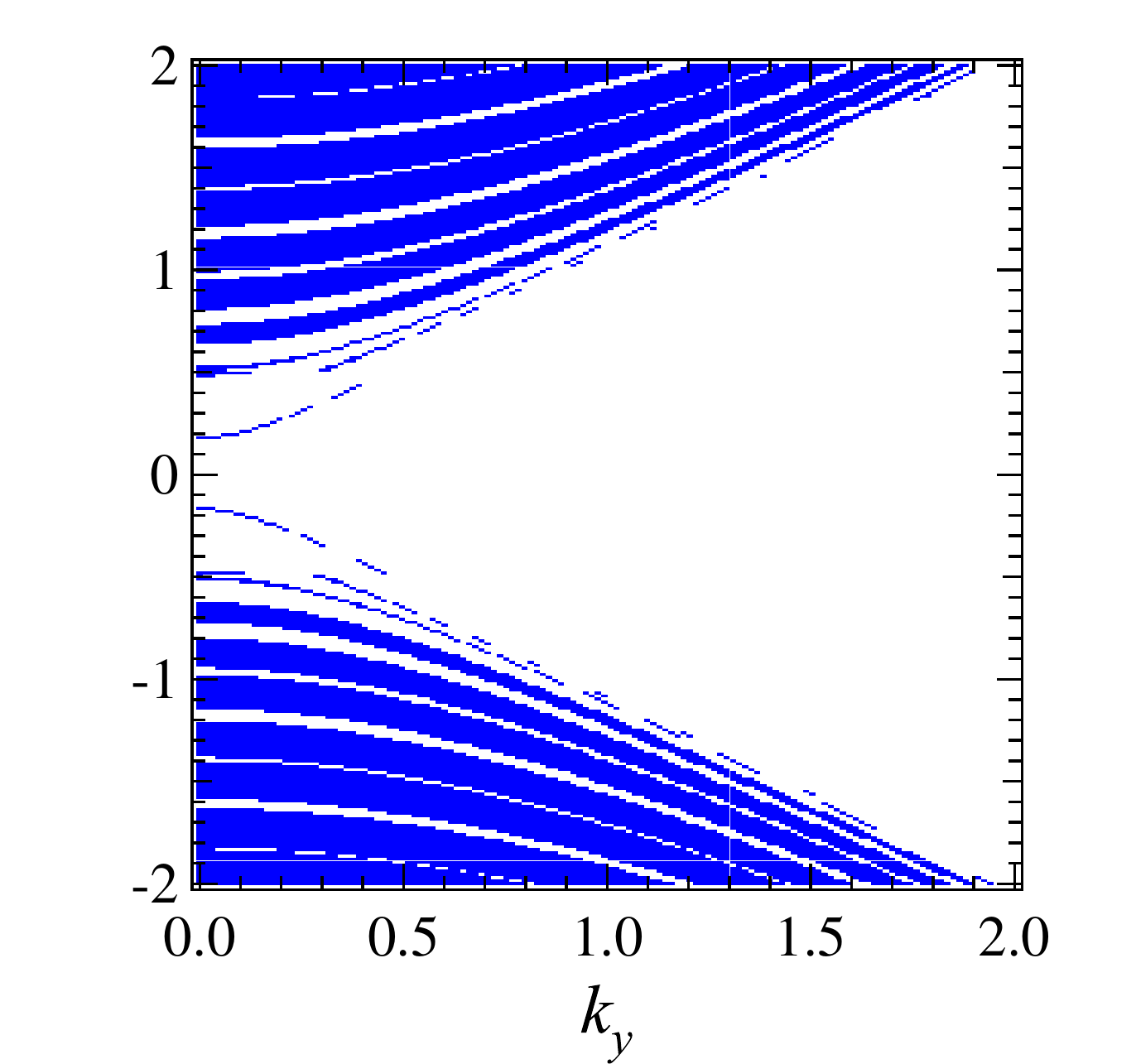}}
\subfigure[][~$\lambda=1/4$ and $\Delta=1/4$]{\label{fig:contourplot-c}\includegraphics[width=0.23\textwidth]{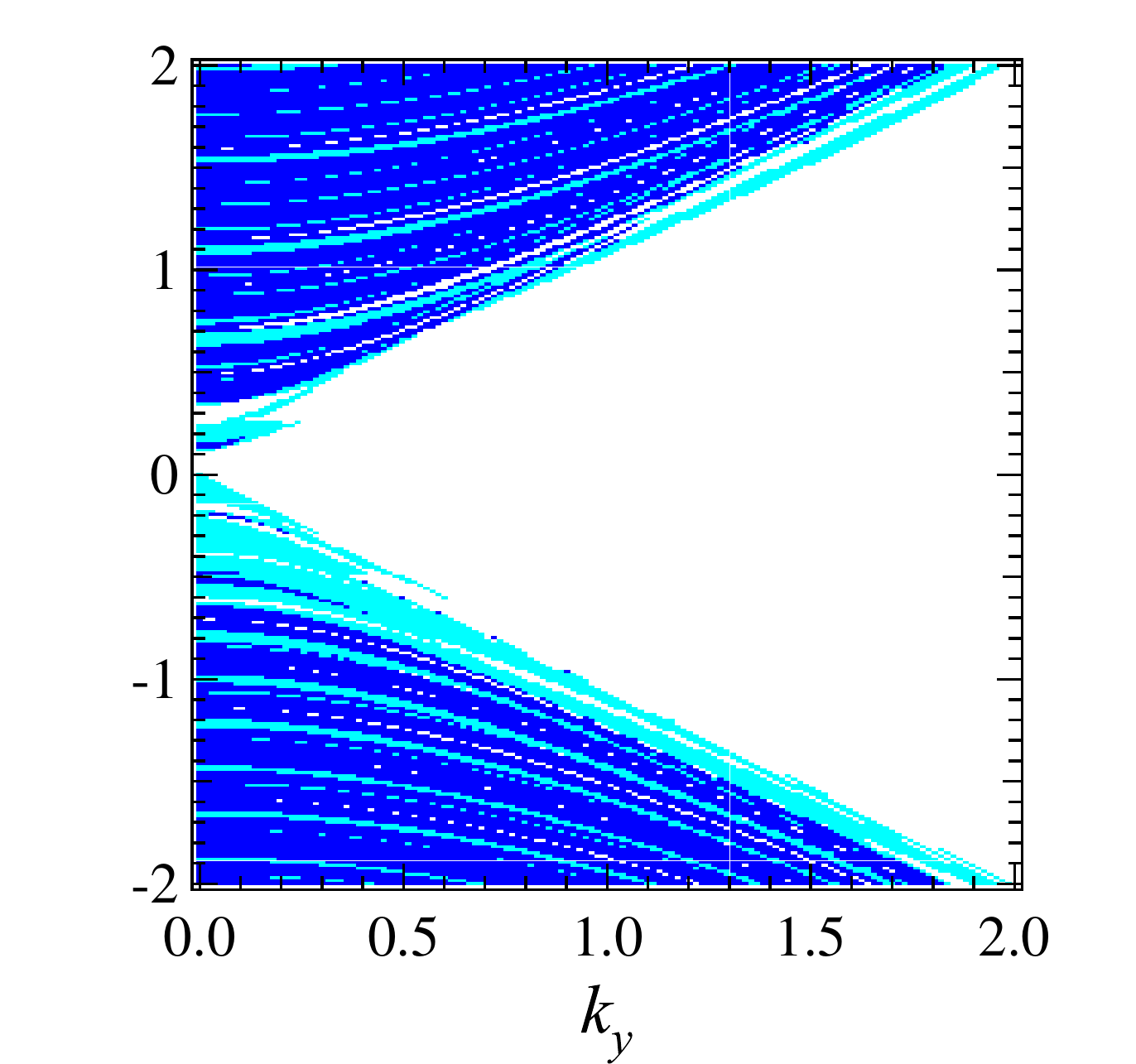}}
\subfigure[][~$\lambda=1/2$ and $\Delta=1/4$]{\label{fig:contourplot-d}\includegraphics[width=0.23\textwidth]{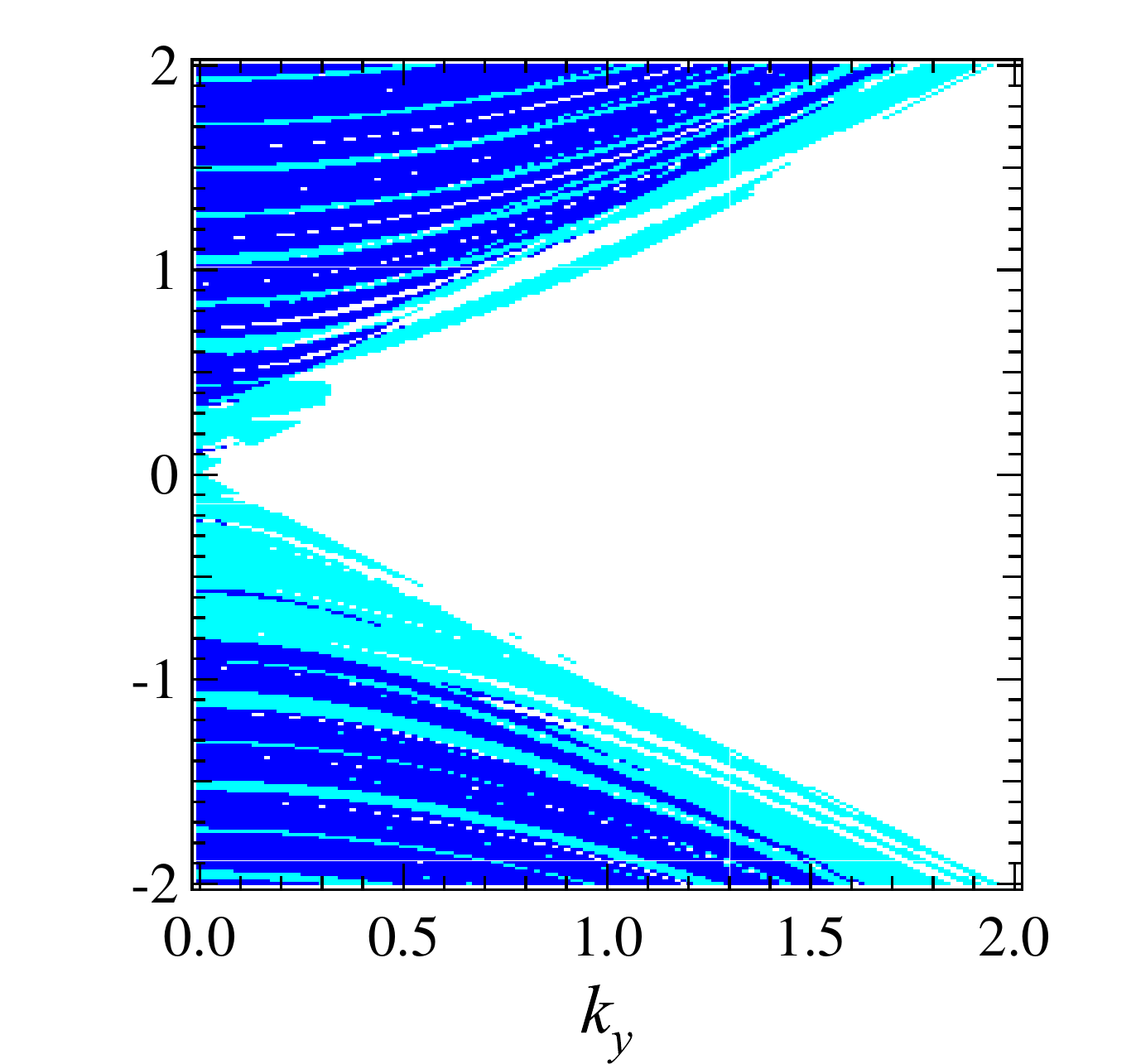}}
\caption{The figures show the number of allowed bands as a function of $k_y$ and $\varepsilon$. Blue corresponds to four allowed bands, turquoise to two and white to zero. The SOI potential length is $a=7$ and the length without the potential is $b=7$. }
\label{fig:contourplot}
\end{figure*}
%
%

\section{Finite propagation angle}
After a rather complete discussion of the band structures with $\phi=0$ we  turn to the case $\phi\not=0$. As explained before, for this case Eq.~\eqref{eq:bandfinal} is not valid. Hence, for finite angles $\phi$ we evoke Bloch's theorem, which states that  
%
%
\begin{align}
 \bm{\psi}(x+\ell)=\ee^{\ii \kappa \ell}\ \bm {\psi}(x).
\label{eq:bloch}
\end{align}
Thus the only change of the spinor after one unit cell is the multiplication by a phase factor $\ee^{\ii\kappa \ell}$ | therefore
%
%
%
%
\begin{equation}
\bm{\mathcal{C}}_{n+1}=
\ee^{\ii\kappa \ell} \, \mathbb{I}_4\cdot
\bm{\mathcal{C}}_n = \mathcal{B}\cdot \bm{\mathcal{C}}_n
\label{eq:transfer2}
\end{equation}
%
%
where $\mathbb{I}_4$ is the $4\times4$ identity matrix. Now, equations \eqref{eq:transfer1} and \eqref{eq:transfer2} can be combined in in the following equation
%
%
\begin{align}\label{eq:bandcon}
 (\mathcal{T-B}) \cdot \bm{\mathcal{C}}_n\
=0\,.
\end{align}
%
%
which is the band condition for finite $\phi$. Note that in absence of SOIs the method summarized by Eq.~\eqref{eq:bandcon} is equivalent to the TM method~\cite{mckellar:1987}. But when SOIs are considered and a four-dimensional space is needed, it gives all conditions on the eigenvalues of $\mathcal{T}$ while the TM method gives only conditions on the sum of eigenvalues.

Figure~\ref{fig:contourplot} shows the number of solutions of the band equation~\eqref{eq:bandcon} as a function of $\varepsilon$ and $k_y$ for all the allowed values of $\kappa$ in the first Brillouin zone.
As in the previous cases, these results can be explained by looking at the dispersion relation for uniform SOI. There, a finite value of $k_y$ | equivalent to a finite value of $\phi$ |  leads to an energy gap at $k_x=0$ even if $\Delta=0$.  Accordingly, in panel \subref{fig:contourplot-b}, where the number of allowed bands for the modulated case with $\lambda=1$ and $\Delta=0$ is shown, a band gap growing with $k_y$ is displayed. Another feature, which is the same for the uniform SOI case and the present one, is that if either  of $\lambda$ and $\Delta$ are zero, electron-hole symmetry is also preserved for finite $k_y$.
This can be seen, although only in terms of the number of allowed bands, in panel \subref{fig:contourplot-b} and in panel \subref{fig:contourplot-a}, where either of $\Delta$ and $\lambda$ are zero and the electron-hole symmetry is preserved, while this is not the case in panel \subref{fig:contourplot-c} and \subref{fig:contourplot-d}, where both, $\lambda$ and $\Delta$, are finite. 
\section{Summary}
We have studied the band structure of graphene under a periodically modulated SOI. In this case spin-full fermions have to be considered, transforming the original SU(2) structure of the graphene Hamiltonian into a SU(2)$\otimes$SU(2) superstructure. For this case the TM method | commonly used for calculating band structures | is not applicable. 
Nevertheless, we have proposed a generalization of the TM method for the spin-dependent case at $\phi=0$. The new method is possible because of the special form of the matrix of spinors inside the SO potential.
This permits us to calculate analytic equations [c.f.~Eq.~\eqref{eq:bandfinal}] for the spin-polarized bands, which in turn  enables us to estimate the gaps due to avoided band crossings at the Brillouin zone boundaries [c.f.~Eq.~\eqref{eq:gaps}]  via a perturbative approach. In addition, we  analyze the dependence of the band structure on the length of the SOI potential compared to the spin-precession length [c.f.~Fig.~\ref{fig:spinprec}]. Here we consider the spin precession length in the limit of energies much larger than the SOI coupling constants, where it is independent of the energy [c.f.~Eq.~\eqref{eq:spinprec}].  Furthermore, we investigate the band structures for finite $\phi$ within an approach based on the Bloch's theorem. This leads to a band equation [c.f.~Eq.~\eqref{eq:bandcon}], which we solve numerically. We find that the modification to the band structure due to finite angles $\phi$ can be understood by considering the properties of the energy dispersion for the case of uniform SOI. 

\acknowledgments
We acknowledge A.~De Martino, H.~Grabert, P.A.M.~Schijven and D.F.~Urban for useful discussions. Our work is supported by the DFG grant BE~4564/1-1 and by the Excellence Initiative of the German Federal and State Governments.


\begin{thebibliography}{0}
\bibitem{reviews} 
A.~K. Geim and K.~S. Novoselov,
Nature Mat. {\bf 6}, 183 (2007); A.H. Castro Neto \emph{et al.}, 
Rev. Mod. Phys. {\bf 81}, 109 (2009); T. Ando, J. Phys. Soc. Jpn. \textbf{74}, 777 (2005).

\bibitem{mckellar:1987} 
B.~H.~J. McKellar and G.~J. Stephenson, Jr.,  Phys. Rev. C \textbf{35}, 2262 (1987).

\bibitem{park:2008} C.-H. Park, \emph{et al.} Nat. Phys. \textbf{4}, 213 (2008).

\bibitem{Barbier:2008}
M. Barbier, F.~M. Peeters, P. Vasilopoulos, and J.~M. Pereira,
Phys. Rev. B \textbf{77}, 115446 (2008).

\bibitem{Barbier:2009} M. Barbier, P. Vasilopoulos, and F.~M. Peeters, Phys. Rev. B \textbf{80}, 205415 (2009); Phys. Rev. B \textbf{81}, 075438 (2010).

\bibitem{dellanna:2010} L. Dell'Anna and  A. De Martino, Phys. Rev. B \textbf{83}, 155449 (2011) and references therein.

\bibitem{kane:2005} 
C.~L. Kane and E.~J. Mele, 
Phys. Rev. Lett. {\bf 95}, 226801 (2005).

\bibitem{soguinea}
D. Huertas-Hernando, F. Guinea, and A. Brataas,
Phys. Rev B {\bf 74} 155426 (2006).

\bibitem{so2}
Min Hongki, \emph{et al.}, 
Phys. Rev. B {\bf 74}, 165310 (2006).

\bibitem{so3}
Y. Yao, F. Ye, X.~L. Qi, S.~C. Zhang, and Z. Fang,
Phys. Rev. B {\bf 75}, 041401(R) (2007).

\bibitem{so4} 
J.~C. Boettger and S.~B. Trickey, Phys. Rev. B \textbf{75}, 121402(R)
(2007); Phys. Rev. B \textbf{75}, 199903(E) (2007).

\bibitem{weeks:2011} C. Weeks \emph{et al.}, Phys. Rev. X \textbf{1}, 021001 (2011).

\bibitem{so5} 
M. Zarea and N. Sandler, Phys. Rev. B \textbf{79}, 165442 (2009).

\bibitem{Konschuh:2010} S.~Konschuh, M.~Gmitra, and J.~Fabian, Phys. Rev. B \textbf{82}, 245412 (2010). 

\bibitem{varykhalov:2008} 
A. Varykhalov \emph{et al.}, 
Phys. Rev. Lett. \textbf{101}, 157601 (2008). 

\bibitem{PhysRevLett.102.057602} O. Rader \emph{et at.}, Phys. Rev. Lett. \textbf{102}, 057602 (2009).

\bibitem{varykhalov:2009} A. Varykhalov and O. Rader, 
Phys. Rev. B \textbf{80}, 035437 (2009).

\bibitem{spintronics}I. \v Zuti\'c, J. Fabian, S. Das Sarma,
Rev. Mod. Phys. {\bf 76}, 323 (2004).

\bibitem{spin:pumping} D. J. Thouless, Phys. Rev. B, \textbf{27} (1983) 6083; V. K. Dugaev, E. Ya. Sherman, and J. Barna\'s Phys. Rev. B \textbf{83}, 085306 (2011); F. S. M. Guimar\~aes, A. T. Costa, R. B. Muniz, and M. S. Ferreira,  Phys. Rev. B \textbf{81}, 233402 (2010). 

\bibitem{Reimann:1997} P. Reimann, M. Grifoni, and P. H\"anggi, Phys. Rev. Lett. \textbf{79}, 10 (1997). 

\bibitem{smirnov:2008} S. Smirnov, D. Bercioux, M. Grifoni, and K. Richter, Phys. Rev. Lett. \textbf{100}, 230601 (2008).

\bibitem{scheid:2007} M. Scheid, D. Bercioux, and K. Richter, New J. Phys. \textbf{9}, 401 (2007). 

\bibitem{rashbagraphene} E.~I. Rashba, Phys. Rev. B {\bf 79}, 161409(R) (2009).

\bibitem{bercioux:2010} D. Bercioux and A. De Martino, Phys. Rev. B \textbf{81}, 165410 (2010).

\bibitem{lenz:soon} A proof is possible considering the subblocks of $\Omega_\so$ when doing its inversion and writing it as a sum of two matrices which have two zero eigenvalues each and two eigenvalues multiplying to one. An article containing the proof is in preparation.

\bibitem{ashcroft:book} N.~W. Ashcroft and N.~D. Mermin, \emph{Solid State Physics}, (Saunders College Publishing, 1976).

\end{thebibliography}
\end{document}